\documentclass[prc,twocolumn,floatfix,groupedaddress,nofootinbib,preprintnumbers,
amsmath,amssymb,amsfonts,superscriptaddress,widetable] {revtex4}
\usepackage{bm}
\usepackage{mathrsfs}
\usepackage{amssymb}
\usepackage{amsmath}
\usepackage{graphicx}
\usepackage{array}
\usepackage{color}

\begin{document}
\title{Weak radius of the proton}
\author{C. J. Horowitz}\email{horowit@indiana.edu}
\affiliation{Center for Exploration of Energy and Matter and
                  Department of Physics, Indiana University,
                  Bloomington, IN 47405, USA}
\date{\today}
\begin{abstract}



















The weak charge of the proton determines its coupling to the $Z^0$ boson.  The distribution of weak charge is found to be dramatically different from the distribution of electric charge.  The proton's weak radius $R_W= 1.545\pm 0.017$ fm is 80\% larger than its charge radius $R_{ch}\approx 0.84$ fm because of a very large pion cloud contribution.  This large weak radius can be measured with parity violating electron scattering and may provide insight into the structure of the proton, various radiative corrections, and possible strange quark contributions.

\end{abstract}
\maketitle
The distributions of charge and magnetization provide crucial insight into the structure of the proton.   Hofstadter in the 1950s determined that the charge distribution in the proton has a significant size $\approx 0.8$ fm \cite{hofstadter}, while experiments at Jefferson Laboratory have shown that the magnetization is distributed differently from the charge \cite{JLAB,JLAB2,JLAB3}.  More recently, an experiment with muonic hydrogen has found a surprising value for the charge radius of the proton $R_{ch}$ \cite{MuonicH} that is smaller than previous determinations with electron scattering or conventional hydrogen spectroscopy \cite{CODATA}.  This proton radius puzzle has motivated considerable theoretical and experimental work and is presently unresolved \cite{Puzzle}.        

In addition to its electric charge, the proton has a weak charge $Q_p$ that characterizes the strength of the vector coupling to the $Z^0$ boson.  The Jefferson Laboratory $Q_{\rm weak}$ collaboration has recently measured $Q_p$ using parity violating electron scattering \cite{QWEAK},
\begin{equation}
Q_p= 0.0719\pm0.0045\, .
\label{Qweak}
\end{equation}
This value provides a sensitive test of the standard model at low energies, and the future P2 experiment aims to improve the accuracy further \cite{P2}.

In the standard model, the weak neutral current is a mixture of the isovector weak and electromagnetic currents.  As a result the Sachs form factor that describes the vector interaction of the $Z^0$ boson with the proton, $G_E^{Zp}$ is related to the conventional electric form factors of the proton $G_E^p$ and neutron $G_E^n$ plus a possible strange quark contribution $G_E^s$ \cite{Weak_Review},
\begin{equation}
4G^{Zp}_E(q^2) = Q_pG^p_E(q^2)+Q_nG^n_E(q^2)-G^s_E(q^2).
\label{GZp}
\end{equation}
Here $q^2=-q_\mu^2>0$ is the square of the momentum transfer and $Q_n$ is the weak charge of the neutron.  


The distribution of weak  charge in the proton is characterized by a size or root mean square radius $R_W$ that follows from the $q^2$ dependence of the form factor $4G^{Zp}_E(q^2)$.   Using
\begin{equation}
Q_p R_W^2=-6\frac{d}{dq^2} (4G_E^{Zp})\Bigl|_{q^2=0},
\label{eq.Rwnorm}
\end{equation}
one has,
\begin{equation}
R_W^2=R_{ch}^2+\frac{Q_n}{Q_p}R_n^2 -\frac{1}{Q_p}R_s^2\, .
\label{R_W}
\end{equation}
Here the neutron charge radius squared is $R_n^2=-0.1148\pm 0.0035$ fm$^2$ \cite{Rn2}.  Note that this $R_n^2$ value is a somewhat old result that could be improved with a modern experiment.   The nucleon's strangeness radius squared $R_s^2$ is defined,
\begin{equation}
R_s^2=-6\frac{d}{dq^2}G_E^s\bigl|_{q^2=0}\, .
\end{equation}  
There have been several measurements of $G_E^{Zp}$ and $G^s_E$ using parity violating electron scattering, see for example \cite{HappexHe, Global}.  Recently the $Q_{\rm weak}$ collaboration determined $R_s^2$ in addition to $Q_p$ \cite{QWEAK} (from their $\rho_s$ value),
\begin{equation}
R_s^2=-0.013\pm 0.007\, {\rm fm}^2.
\label{eq.RsQw}
\end{equation}
Alternatively, Lattice QCD calculations \cite{LQCD} find a consistent but more precise value,
\begin{equation}
R_s^2=-0.0054\pm 0.0016\, {\rm fm}^2\, \ \  ({\rm LQCD}),
\label{eq.RsLQCD}
\end{equation}
where we have combined their different errors in quadrature.  If the $Q_{\rm weak}$ collaboration constrain $R_s^2$ to the LQCD value, they find a slightly different value for $Q_p$ instead of Eq. \ref{Qweak},
\begin{equation}
Q_p= 0.0685\pm0.0038\, .
\label{QLQCD}
\end{equation}
We note that Eq.~\ref{QLQCD} is consistent with Eq.~\ref{Qweak} within errors and both are consistent with the standard model value, see below.

We list in Table \ref{Table1} values of the proton weak radius from Eq.~\ref{R_W}.  These involve the large ratio of the neutron to proton weak charges.  At tree level $Q_n=-1$ and $Q_p=1-4\sin^2\Theta_W\approx 0.05$.  Including radiative corrections yields the standard model values $Q_n=-0.9902$ and $Q_p=0.0710$ \cite{PDG}\footnote{We neglect radiative corrections for the strange quark contribution.}.  Our procedure is to use the LQCD value for $R_s^2$ and the standard model values for $Q_p$ and $Q_n$.  This yields the most precise $R_W$ where the error is dominated by the error in $R_n^2$.    If we use the $Q_{\rm weak}$ value for $Q_p$ from Eq.~\ref{QLQCD} instead, we have a similar value for $R_W$ with a slightly larger $\approx\pm 0.035$ fm error.  Table \ref{Table1} has two lines for $R_W$ because of the proton radius puzzle.  The second line (and the line for the neutron, see below) corresponds to the smaller $R_{ch}$ from muonic hydrogen.  
\begin{table}[ht]
\begin{tabular}{cccc}
Particle                 & $R_{ch}$ (fm)           & $R_W$ (fm)        & $\Delta R$ (fm)  \\

 \hline
p & $0.877\pm0.007$ \cite{CODATA}& $1.564\pm0.017$ & $0.687\pm 0.017$ \\
 & $0.8418\pm0.0007$ \cite{MuonicH}& $1.545\pm 0.017$ & $0.703 \pm 0.017$ \\
 n & & $0.8434\pm0.0012$& \\ 
 $^{208}$Pb & 5.503 & $5.826\pm0.181$\cite{PREX} & $0.323\pm0.181$\\ 
\end{tabular}
\caption{The charge radius $R_{ch}$, weak radius $R_W$, and weak skin $\Delta R=R_W-R_{ch}$ for the proton, neutron,  and $^{208}$Pb nucleus. }
\label{Table1}
\end{table}

As shown in Table \ref{Table1}, the weak radius of the proton $R_W$ is $\approx 80$\% larger than the charge radius $R_{ch}$.  We define the weak skin $\Delta R$ as the difference between the weak and charge radii $\Delta R=R_W-R_{ch}$.  The large value $\Delta R\approx 0.7$ fm shows that weak charges are more likely to be found at large distances from the origin than are electric charges.

The dramatic difference $R_W\gg R_{ch}$ is a major result of this paper.  Why does the distribution of weak charge have a spatial extent that is much greater than the extent of the (electric) charge?  We first present an explanation in terms of hadronic coordinates and then we present an alternative description in terms of quark coordinates.  Consider a virtual transition $p\rightarrow n + \pi^+$.  The weak charge of the pion $Q_{\pi^+}=Q_p-Q_n=1.061$ is much larger than $Q_p$.  Therefore the pion ``tail", present in the proton at large radius, ``wags the dog" and makes a very large contribution to $R_W$.

The weak radius of the proton is related to the charge radius (squared) of the neutron.  Consider a virtual transition $n\rightarrow p + \pi^-$.  The $\pi^-$ produces a negative charge distribution at large distances resulting in a negative $R_n^2$.  Thus $R_n^2$ is negative because of the neutron's pion cloud.

Another equivalent way to understand $R_W\gg R_{ch}$ is to consider the distribution of up and down quarks.  The weak charge of a proton $Q_p$ is small because of a sensitive cancelation between the weak charges of two up quarks and the weak charge of one down quark.  Therefore $\rho_W(r)$ is very sensitive to small differences between the distributions of up and down quarks.  If the up quarks have a somewhat larger radius than the down quarks this will lead to $R_W\gg R_{ch}$.  

We note that $R_W$ is sensitive to radiative corrections.  If one evaluates $R_W$ in Eq.~\ref{R_W} with the tree level weak charges $Q_n=-1$ and $Q_p=0.05$ the result is $R_W\approx 1.8$ fm.  This is about 15\% larger than the values in Table \ref{Table1}.  One should study further the impact of radiative corrections on $R_W$.

For completeness, we also discuss the weak magnetic radius of the proton.  The form factor describing the magnetic coupling of the $Z^0$ to the proton $G^{Zp}_M$ has the same form as Eq.~\ref{GZp},
\begin{equation}
4G^{Zp}_M(q^2)=Q_pG_M^p(q^2)+Q_nG^n_M(q^2)-G^s_M(q^2)\, .
\end{equation}
Since $|Q_n|\gg Q_p$, we expect the weak magnetic radius of the proton to be close to the (conventional) magnetic radius of the neutron, and very different from $R_W$.

\begin{figure}[ht]
\includegraphics[width=1.0\columnwidth]{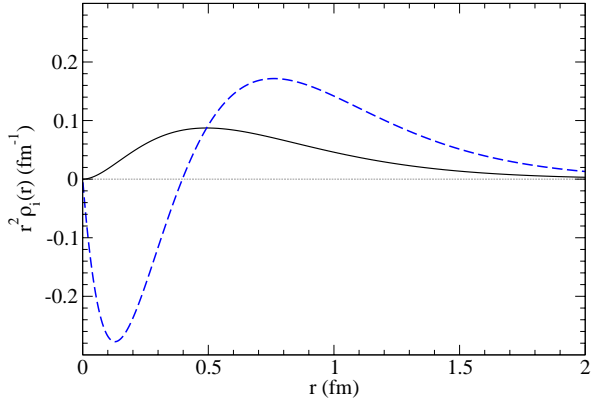}
 \caption{(Color online) Fourier transforms of the proton electric form factor $r^2\rho(r)$ (solid black line) and the normalized weak form factor  $r^2\rho_{W}(r)/Q_p$ (dashed blue line) versus radius $r$, see Eq.~\ref{rho3W}.  
 }
\label{Fig1}
\end{figure}

To gain additional insight into $R_W\gg R_{ch}$, we consider coordinate space representations related to the local charge or weak charge density.  We define the function $\rho_{W}(r)$ as the three dimensional Fourier transform of $G^{Zp}_E(q^2)$,
\begin{equation}
\rho_{W}(r)=\int\frac{d^3q}{(2\pi)^3}\, {\rm e}^{-i{\bf q\cdot r}} 4G^{Zp}_E(q^2)\, .
\label{rho3W}
\end{equation}
In the nonrelativistic limit $\rho_{W}(r)$ corresponds to the weak charge density of the proton.  However relativistic effects can complicate this interpretation.  We will consider a relativistic transverse density below \cite{Miller}.  For now $\rho_{W}(r)$ is defined by Eq.~\ref{rho3W}.  We assume simple form factors $G_E^p(q^2)=(1+q^2/\Lambda^2)^{-2}$ and  $G_E^n(q^2)=aq^2(1+q^2/\Lambda^2)^{-2}$ with the constants $\Lambda=\sqrt{12}/R_{ch}$ and $a=-R_n^2/6$ chosen to reproduce mean square radii.  Of course more detailed form factors can be used, see for example \cite{Ye:2017gyb,Venkat:2010by,Kelly:2004hm}, but we do not expect them to make qualitative differences.   We neglect small strange quark contributions.  This gives a simple analytic form for $\rho_{W}(r)$,
\begin{equation}
\rho_{W}(r)=\Bigl\{Q_p+Q_n\frac{R_n^2}{6}\bigl(\Lambda^2-2\frac{\Lambda}{r}\bigr)\Bigr\}\frac{\Lambda^3}{8\pi}{\rm e}^{-\Lambda r}\, .
\label{rho_W(r)}
\end{equation}     
In Fig.~\ref{Fig1} we plot the normalized function $\rho_{W}(r)/Q_p$.  We see that $\rho_{W}(r)$ has a node near $r=0.4$ fm.  

For reference, Fig. \ref{Fig1} also shows the function $\rho(r)=\Lambda^3{\rm e}^{-\Lambda r}/(8\pi)$ determined from the Fourier transform of the proton electric form factor $G_E^p(q^2)$.  There are large differences between $\rho_{W}(r)$ and $\rho(r)$.  Note that the functions in Fig. \ref{Fig1} have been multiplied by $r^2$ to emphasize their large $r$ behaviors.

We now consider a relativistically consistent representation of the weak charge density of the proton provided by the transverse density $\rho_W^t(b)$ \cite{Miller}.  This is the two dimensional Fourier transform of the weak $F_1^{Zp}$ form factor,
\begin{equation}
\rho_W^t(b)=\int \frac{d^2q}{(2\pi)^2}{\rm e}^{-i{\bf q\cdot b}} 4F_1^{Zp}(q^2)\, ,
\label{rhoT}
\end{equation}
\begin{equation}
\rho_W^t(b)=\int \frac{qdq}{2\pi} J_0(qb) \frac{4G_E^{Zp}+\tau 4G_M^{Zp}}{1+\tau}\, .
\label{rhoT2}
\end{equation}
Here $\tau=q^2/4M_N^2$ and $J_0$ is a cylindrical Bessel function.  We neglect strange quark contributions and use simple parameterizations of the form factors $G_E^n=\mu_n\tau G_E^p/(1+5.6\tau)$ \cite{Glaster}, $G_M^n=\mu_n G_E^p$, and $G_M^p=(1+\mu_p)G_E^p$ with $\mu_p=1.7928$, $\mu_n=-1.9130$ and $G_E^p=(1+q^2/\Lambda^2)^{-2}$.  The impact of more accurate fits, see for example \cite{FF}, should be explored.

\begin{figure}[ht]
\includegraphics[width=1.0\columnwidth]{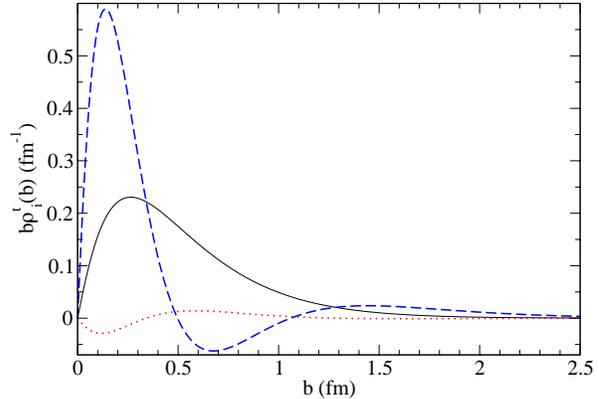}
 \caption{(Color online) Impact parameter $b$ times the transverse charge density of the proton $b\rho^t(b)$ (solid black line), transverse weak density of the proton $b \rho_W^t(b)/Q_p$ (dashed blue line) and the transverse charge density of the neutron $b \rho_n^t(b)$ (dotted red line), see Eq.~\ref{rhoT}.
 }
\label{Fig2}
\end{figure}

Figure \ref{Fig2} shows the transverse weak density $\rho_W^t(b)/Q_p$, and the transverse charge densities of the proton $\rho^t(b)$, and neutron $\rho_n^t(b)$.  These are defined as in Eq.~\ref{rhoT} using $F_1^p$ and $F_1^n$.  
At large distances both $\rho_W^t(b)/Q_p$ and $\rho_W(r)/Q_p$ are large and positive.  This shows the large contribution of the pion tail.   Furthermore, $\rho_W^t(b)/Q_p$ is very different from $\rho^t(b)$.  Likewise $\rho_W(r)/Q_p$ is very different from $\rho(r)$.  However $\rho_W^t(b)$ is positive for small impact parameters while $\rho_W(r)$ is negative at small $r$.    Note that small impact parameter $b$ does not necessarily correspond to small radius $r$.  The large differences seen in Figs. \ref{Fig1}, \ref{Fig2} emphasize that the weak charge in a proton is distributed very differently from the electric charge.

We summarize relativistic effects.  In general, all of the equations in this paper are fully relativistic and there are no relativistic corrections to any of our numerical results.  The only complication from relativity is in the physical interpretation of $\rho_W(r)$ that is defined by the three dimensional Fourier transform in Eq.~\ref{rho3W}.  In order to probe the system at small distances, one must scatter with a momentum transfer $q$ that is comparable to the nucleon's mass.  This causes the nucleon to recoil and this recoil must be taken into account.  Therefore $\rho_W(r)$ shown in Fig.~\ref{Fig1} can not be interpreted as a local weak charge density in coordinate space.  In contrast the transverse density  $\rho_W^t(b)$ defined in Eq.~\ref{rhoT} and shown in Fig.~\ref{Fig2} can be interpreted as a local weak charge density because it is not impacted by boosting the system along the direction of $q$ from the nucleon's recoil.


For comparison, we now discuss weak radii for the neutron and heavier nuclei.  The form factor describing the coupling of the $Z^0$ to the neutron is $G^{Zn}_E$,
\begin{equation}
4G^{Zn}_E(q^2) = Q_nG^p_E(q^2)+Q_pG^n_E(q^2)-G^s_E(q^2) .
\label{GZn}
\end{equation}
Defining the neutron weak radius from $Q_n{R_W^n}^2=-6d(4G^{Zn}_E)/dq^2|_{q^2=0}$ yields,
\begin{equation}
{R_W^n}^2=R_{ch}^2+\frac{Q_p}{Q_n}R_n^2-\frac{1}{Q_n}R_s^2\, .
\end{equation}
Given $|Q_n|\gg Q_p$, the weak radius of the neutron, see Table \ref{Table1}, is very close to the charge radius of the proton $R^n_W\approx R_{ch}$.  For light $N=Z$ nuclei such as the deuteron, $^4$He or $^{12}$C the weak radius of a nucleus is expected to be close to its charge radius.  For mirror nuclei such as $^3$He and $^3$H the weak radius of $^3$He should be close to the charge radius of $^3$H and vise versa.   

Can one gain insight by comparing the weak skin of the proton to the weak skin of heavy nuclei?  We note that Hofstadter explored the charge densities of the proton and heavy nuclei with very similar experiments.  For heavy nuclei with $N>Z$ we expect a neutron skin with some of the extra neutrons collecting in the surface region so that the neutron radius is greater than the proton radius.  As a result there will be a weak skin with $R_W>R_{ch}$.  This has now been verified for $^{208}$Pb, where $R_W$ has been measured in the PREX experiment \cite{PREX,PREX1}, see Table \ref{Table1}.  

Clearly the weak skin of the proton is not produced by a neutron skin even though both the proton and $^{208}$Pb have $R_W>R_{ch}$.  Instead, the weak skin of the proton can be thought of as coming from an ``up quark skin'' rather than a neutron skin.  The up quark skin describes an excess of up quarks at large radii in the proton.  For example, a virtual $\pi^+$ cloud at large radii will increase the density of up quarks and reduce the net density of down quarks (minus down antiquarks) and produce an up quark skin.  Thus, the proton could also be thought of as having a ``pion skin''.

We end by exploring if there are nuclei with $R_W\gg R_{ch}$ that might have weak skins at least somewhat comparable to the very large $\Delta R\approx 0.7$ fm of the proton.  We consider two possibilities.  The first is a neutron halo nucleus such as $^{11}$Li where the neutron radius and $R_W$ may be dominated by the weakly bound neutron halo that extends to very large radii \cite{Li11}.  The second possibility is simply a very neutron rich nucleus.  This nucleus will likely have a thick neutron skin (and hence weak skin) because of all of the ``extra'' neutrons.  

It is interesting to consider Ca isotopes.   The heaviest stable $N=Z$ nucleus is $^{40}$Ca where we expect $R_{ch}$ to be slightly larger than $R_W$ ($\Delta R<0$) because the protons are pushed out by the Coulomb interaction.   The CREX experiment should accurately measure $R_W$ for the doubly closed shell neutron rich isotope $^{48}$Ca \cite{CREX}.  Here $R_W$ depends on poorly constrained three neutron forces.  Microscopic chiral effective field theory calculations predict a thin weak skin for $^{48}$Ca, $\Delta R\approx 0.13$ fm \cite{CC48Ca} while dispersive optical model calculations predict a thick skin $\Delta R\approx 0.25$ fm \cite{DOM48Ca}.  Even more neutron rich Ca isotopes are expected to have thicker weak skins.  Recently the isotopes $^{59}$Ca and $^{60}$Ca were observed at RIKEN \cite{Ca60} suggesting that very neutron rich Ca isotopes, perhaps up to $^{70}$Ca, are particle stable.  The nucleus $^{70}$Ca, with 2.5 times more neutrons than protons, could have a very thick weak skin.


In conclusion, we have calculated the distribution of weak charge in the proton and find it to be dramatically different from the distribution of electric charge.  The weak radius $R_W$ is $\approx 80$\% larger than the charge radius $R_{ch}$ because of a very large pion cloud contribution.  This large weak radius probes proton structure including differences between up and down quark distributions, and $R_W$ can be measured with parity violating electron scattering.

\begin{acknowledgments}
We thank Bill Donnelly, Farrukh Fattoyev, Misha Gorshteyn, Zidu Lin, Jerry Miller, Seamus Riordan, and Mike Snow for helpful discussions.  This work was started at the Mainz Institute for Theoretical Physics and we thank them for their hospitality.  This material is based upon work supported by the U.S. Department of Energy Office of Science, Office of Nuclear Physics under Awards DE-FG02-87ER40365 (Indiana University) and DE-SC0018083 (NUCLEI SciDAC-4 Collaboration).
\end{acknowledgments}


\end{document}